%% file: main.tex
\documentclass[sigconf]{acmart}

\copyrightyear{2026}
\acmYear{2026}
\setcopyright{cc}
\setcctype{by}
\acmConference[ICSE '26]{2026 IEEE/ACM 48th International Conference on Software Engineering}{April 12--18, 2026}{Rio de Janeiro, Brazil}
\acmBooktitle{2026 IEEE/ACM 48th International Conference on Software Engineering (ICSE '26), April 12--18, 2026, Rio de Janeiro, Brazil}
\acmPrice{}
\acmDOI{10.1145/3744916.3773249}
\acmISBN{979-8-4007-2025-3/26/04}

\usepackage{amsmath,amsfonts}
\usepackage{algorithmic}
\usepackage{graphicx}
\usepackage{textcomp}
\usepackage{enumitem}
\usepackage{xcolor}
\usepackage{xspace}
\usepackage{booktabs}
\usepackage[most]{tcolorbox}
\usepackage{subfigure}
\usepackage{multirow}
\usepackage{algorithm}
\usepackage{colortbl}
\usepackage{url}
\usepackage{booktabs}
\newlength{\thinline}
\newlength{\thickline}
\definecolor{mygray}{gray}{0.9}
\definecolor{deepgray}{gray}{.8}
\newcommand{\xm}[1]{\textcolor{black}{#1}}
\newcommand{\wht}[1]{\textcolor{black}{#1}}

\newcommand{\name}{\textit{RulePilot}\xspace}

\usepackage{tabularx}
\usepackage{varwidth}
\def\BibTeX{{\rm B\kern-.05em{\sc i\kern-.025em b}\kern-.08em
    T\kern-.1667em\lower.7ex\hbox{E}\kern-.125emX}}

\begin{document}


\title{\name: An LLM-Powered Agent for Security Rule Generation}

\author{Hongtai Wang\textsuperscript{*}}
\affiliation{%
  \institution{National University of Singapore}
  \city{Singapore}
  \country{Singapore}
}
\email{wanghongtai0702@gmail.com}

\author{Ming Xu\textsuperscript{*}\textsuperscript{\#}}
\affiliation{%
  \institution{National University of Singapore}
  \city{Singapore}
  \country{Singapore}
}
\email{mingxu@nus.edu.sg}

\author{Yanpei Guo}
\affiliation{%
  \institution{National University of Singapore}
  \city{Singapore}
  \country{Singapore}
}
\email{guo.yanpei@u.nus.edu}

\author{Weili Han}
\affiliation{%
  \institution{Fudan University}
  \city{Shanghai}
  \country{China}
}
\email{wlhan@fudan.edu.cn}

\author{Hoon Wei Lim}
\affiliation{%
  \institution{Cyber Special Ops-R\&D, NCS Group}
  \city{Singapore}
  \country{Singapore}
}
\email{hoonwei.lim@ncs.com.sg}

\author{Jin Song Dong}
\affiliation{%
  \institution{National University of Singapore}
  \city{Singapore}
  \country{Singapore}
}
\email{dcsdjs@nus.edu.sg}

\thanks{\textsuperscript{*} Both authors contributed equally to the paper, ordered alphabetically.}
\thanks{\textsuperscript{\#} Corresponding author.}

\begin{abstract} 
The real-time demand for system security leads to the detection rules becoming an integral part of the intrusion detection life-cycle. Rule-based detection often identifies malicious logs based on the predefined grammar logic, requiring experts with deep domain knowledge for rule generation. Therefore, automation of rule generation can result in significant time savings and ease the burden of rule-related tasks on security engineers. 
In this paper, we propose \name, which \wht{mimics human expertise via LLM-based agent for addressing rule-related challenges like rule creation or conversion.}
Using \name, the security analysts do not need to write down the rules following the grammar, instead, they can just provide the annotations such as the \wht{natural-language-based} descriptions of a rule, 
our \name can automatically generate the detection rules without more intervention. 
\name is equipped with the intermediate representation (IR), which abstracts the complexity of config rules into structured, standardized formats, allowing LLMs to focus on generation rules in a more manageable and consistent way. 
We present a comprehensive evaluation of \name in terms of textual similarity and \wht{execution success} abilities, showcasing \name can generate high-fidelity rules, outperforming the baseline models by up to 107.4\% in textual similarity to ground truths and achieving better detection accuracy in real-world execution tests. 
We perform a case study from our industry collaborators in Singapore, showcasing that \name significantly help junior analysts/general users in the rule creation process. 

\end{abstract}



\maketitle

\section{Introduction} 
Security threats are increasingly a growing concern for both users and industrial organizations. The infamous SolarWinds attack~\cite{SolarWinds} disrupted supply chains and compromised sensitive data, highlighting the critical need for robust security controls. A recent trend in intrusion detection systems relies on the neural-network-based provenance graphs, which have demonstrated notable strength in detection performance. However, they face the problems of high computational resource cost and long detection latency, hindering their wide practical deployment.
In practice, in security detection systems, rules~\cite{DBLP:journals/ieeesp/BhattMZ14:SIEM} are widely used to identify malicious activities and trigger alerts, such as detection rules executed on SIEM (Security Information and Event Management) platforms, which offer a lightweight and efficient solution to these challenges while maintaining great explanation abilities.

However, the high cost of rule creation and the long duration of rule maintenance are still problems faced by security organizations. Particularly, these detection rules are typically written manually by junior and senior security experts, a process that is time-consuming, labor-intensive and requires extensive domain knowledge. Furthermore, as attack techniques continue to evolve, rules need constant updates, increasing maintenance costs. 
Tools like MITRE ATT\&CK~\cite{mitre-attack} provide a common framework to describe attack techniques, but translating the structured techniques into specific rule configurations requires huge manual efforts. 
Moreover, modern security organizations can sometimes use different SIEM platforms such as Splunk~\cite{splunk}, Microsoft Sentinel~\cite{microsoftMicrosoftSentinel}, or IBM QRadar~\cite{ibmqradar}, which have their own rule languages. Rules written for one platform cannot directly work on another, creating a cross-platform compatibility problem when an organization migrates an SIEM system. The automation of rule generation and conversion can result in significant time savings and ease the burden of rule-related tasks on security engineers.

The recent breakthrough of Large Language Models (LLMs), particularly in code generation~\cite{DBLP:conf/emnlp/FengGTDFGS0LJZ20:codebert,DBLP:conf/emnlp/0034WJH21-codeT5,DBLP:conf/sigsoft/SvyatkovskiyDFS20:IntelliCode,DBLP:journals/tmlr/LiAZMKMMALCLZZW23:lingming-code}, text-to-SQL~\cite{DBLP:journals/corr/abs-2407-15186:Text2SQL} and binary malware analysis~\cite{DBLP:conf/ccs/Xie00X0024:ReSym,DBLP:journals/pacmpl/ZhangYTWK019:BDS-malware} with generative models like the GPT series, open new opportunities for automated security rule generation and conversion. Unfortunately, compared to code/SQL generation, the challenge of generating security configurations lies in the nuanced and dynamic nature inherent in SIEM-specific rules. Code/SQL often follows well-defined syntax and logical structures, while rule configurations are highly domain-specific, system-dependent, and lack standardized formats. The rule configurations require a deep understanding of systems' behaviors and environment, precise tuning of parameters, dependencies, and iterative corrections, which can vary significantly across SIEM systems.
Informally speaking, a junior programmer can write redundant but correct code, however, he/she may struggle to figure out a correct SIEM-specific rule constraint.
Several works~\cite{Tseng2024UsingLT:Peng-liu,DBLP:journals/corr/abs-2407-05194:LLMCloudHunter} explored the generation of simple detection YAML rules like Sigma, falling short in addressing the more complex and functional challenges specific to SIEM systems due to the following challenges.  

\begin{itemize}[fullwidth,itemindent=0em]
     \item \textbf{Non-standardized format:} The rule configurations are highly domain-specific and lack standardized representations. A standalone LLM typically lacks precise knowledge and cannot simulate human thoughts to break down the complex rule generation into smaller pipelines. 
     To address this, we design an intermediate representation (IR) that can serve as a bridge between high-level requirements and low-level configuration file details. An IR abstracts the complexity of rule configurations into a structured, standardized format that captures essential parameters, and relationships, allowing LLMs to focus on generating configurations in a manageable and consistent way. \wht{The designed IR should be capable of handling the SIEM-specific cases like nested operators, vendor-specific syntax, reducing ambiguity and improving accuracy.} 
    \item \textbf{Iterative correction:} The initially generated rules might be semantically and syntactically incorrect, or logically-nonaligned. 
    To resolve this, we introduce the reflection functions, identifying the potential mistakes upon each step, and refining the identified weakness to optimize \wht{the semantic and syntactic gaps.}
    \wht{Beyond that, our reflection supports logical consistency, rule-field coverage, and the live execution viability, enabling the robust and scalable rule generation.}
    \item \textbf{System dependence:} A sound rule should be able to interact with the live SIEM systems while existing LLMs fall short into autonomously and independently use tools like external SIEM vendor's grammar checks, feedback from live SIEM vendor's APIs, or rule-testing frameworks. We integrate the live Splunk~\cite{splunk} SIEM with LLMs, facilitating validation, optimization, and adaptation of configurations across systems and environments.

\end{itemize}

In this paper, for the first time,  we propose \name, which is an LLM-powered agent facilitating a series of practical scenarios on rule-based detection autonomously: 
1) Using \name, security analysts do not need to write rules following a specific grammar. Instead, they can simply provide annotations, such as rule descriptions in natural language. With this input, our \name can automatically generate detection rules without requiring any further intervention. 
Usually, the descriptions can be divided into preconditions like a rule annotations or the attack types provided by experts. 
We tailor our workflow to Splunk SIEM grammars. 
2) Furthermore, when security analysts update or migrate their SIEM systems, they need the conversion function between the different SIEM vendors. \name supports the conversion between Splunk Processing Language (SPL) and Microsoft Sentinel Kusto Query Language (KQL).



We evaluate \name upon objective similarity for textual alignment with ground truth rules~\cite{Splunk-rules} and semantic evaluator for assessing logical and functional correctness. Results show that \name consistently improves both textual similarity and semantic accuracy, outperforming standalone LLMs by up to 107.4\% in textual similarity. 
We conduct a field study by executing the generated rules in a realistic Splunk environment, evaluating their \wht{execution success} in detecting suspicious activities. The results demonstrate that \name successfully captures the majority of suspicious logs by up to 1.00 F1 score, validating its practical applicability in real-world threat detection scenarios.
Our evaluation yields intriguing insights into the capabilities and limitations of LLMs in rule generation. We discover that LLMs show proficiency in understanding high-level threat descriptions and generating corresponding rules, however, we find that LLMs have difficulty in maintaining field mappings and condition handling, \wht{which necessitates human verification for checking the final results, ensuring the generated rule functions correctly}. 
\xm{We perform a case study from our industry partners, and show that \name significantly facilitates the rule generation of junior analysts/general users in terms of time used, rule quality including the syntax validity and logical alignment.}

We summarize our contributions as follows.

\begin{itemize}[left=0em]
    \item We propose a \xm{novel} workflow \name to address SIEM-specific rule generation and conversion challenges. 
    Our designed immediate representation and reflection modular go beyond general IR and reflection mechanisms, effectively covering the SIEM-specific functions and edge cases such as nested operators and logical consistency, making the process more robust and scalable. 
    \item We tailor our \name to Splunk SIEM system, analyzing the grammars and environments specific to Splunk, seamlessly integrating with Splunk for intelligent and  efficient rule execution. To the best of our knowledge, this is the first, end-to-end and real-time agentic framework for SIEM-rule creation.
    \item We conduct a comprehensive evaluation of \name, employing models like GPT-4o, DeepSeek-V3, and LlaMa-3. 
    \name outperforms the baseline models by up to 107.4\% in textual similarity to ground truths and achieves better detection accuracy in real-world Splunk execution tests.
\end{itemize}

\noindent With its structured reasoning and automation capabilities, \name is poised to become an essential tool for security analysts in rule-relevant tasks. We release all the used datasets in the link~\footnote{\url{https://sites.google.com/view/rulepilot/dataset}} and open-source all code in \footnote{\url{https://github.com/LLM4SOC-Topic/RulePilot}}.




\input{1-motivation.tex}
\input{2-approach.tex}
\input{3-evaluation.tex}

\input{4-RQ4.tex}
\input{5-RQ5.tex}
\input{6-discussion.tex}
\bibliographystyle{ACM-Reference-Format}
\bibliography{my}

\end{document}

%% file: 1-motivation.tex
\lstdefinelanguage{yaml}{
    keywords={true,false,null,y,n},
    keywordstyle=\color{red}\bfseries,
    basicstyle=\ttfamily,
    sensitive=false,
    comment=[l]{\#},
    morecomment=[s]{/*}{*/},
    commentstyle=\color{gray}\ttfamily,
    stringstyle=\color{orange}\ttfamily,
    moredelim=[l][\color{green}]{---},
    moredelim=[l][\color{green}]{...},
    moredelim=[l][\color{cyan}]{-},
    moredelim=[l][\color{cyan}]{where},
}
\lstset{ %
  language=yaml,                
  basicstyle=\ttfamily\footnotesize, 
  numbers=none,                   
  numberstyle=\tiny\color{gray},  
  stepnumber=1,                   
  numbersep=5pt,                  
  backgroundcolor=\color{yellow!10},  
  showspaces=false,               
  showstringspaces=false,         
  showtabs=false,                 
  frame=none,                   
  rulecolor=\color{black},        
  tabsize=2,                      
  captionpos=b,                   
  breaklines=true,                
  breakatwhitespace=false,        
  title=\lstname,                 
  keywordstyle=\color{red},      
  commentstyle=\color{dkgreen},   
  stringstyle=\color{mauve},      
  escapeinside={\%*}{*)},         
  morekeywords={*,...}            
}

\section{Background and Motivation}

\subsection{Rule-based Anomaly Detection} 
Modern anomaly detection systems like Security Information and Event Management (SIEM)~\cite{DBLP:journals/ieeesp/BhattMZ14:SIEM, DBLP:journals/corr/abs-2311-10197:SIME-rule-evasion} typically rely on detection rules to identify potential intrusions, which are widely used due to their lightweight overhead and great explanation abilities. 
The widely-used rules can be typically classified into the general Sigma and the SIEM-specific rules. Sigma is a generic and open signature format for SIEM systems, allowing for flexible rules in YAML format that can be translated into multiple SIEM vendors.
Despite their compatibility with any SIEM vendor, Sigma rules primarily rely on single-pattern matching using regular expressions. They lack support for complex queries, such as SQL-style aggregations, and are unable to execute calculations including statistical analysis or time-window-based computations in SIEM vendors.        
These limitations often result in failures to detect sophisticated attacks involving a series of events or cycles. In contrast, SIEM-specific rules can bridge this gap by incorporating customized conditions with conditional statements and leveraging advanced functions. 
Consider a scenario where an attacker attempts to exfiltrate sensitive data by downloading multiple \textit{".zip"} files from a server. A typical Sigma rule detects this behavior through pattern matching in log fields, such as identifying \textit{".zip"} file requests in URI queries, which relies solely on string-based detection, lacking contextual awareness and deeper behavioral analysis.
As a comparison, take the widely-used SIEM vendor  Splunk~\cite{splunk} as an example, a Splunk-specific rule~\cite{Splunk-rules,Splunk-complex} can implement the detection with more advanced functionality shown in Listing~\ref{lst:splunkrule}.
\begin{lstlisting}[caption={A splunk rule for detection of ZIP file downloads.}, label=lst:splunkrule]
index=web_logs 
| search uri_query="*.zip"
| stats count BY src_ip, uri_query, user_agent
| where count > 5 AND user_agent!="Mozilla/5.0 (friendly-bot)"
| eval message="Potential data exfiltration detected: " . src_ip . " downloading " . count . " ZIP files"
| table _time, src_ip, uri_query, user_agent, count, message
\end{lstlisting} 
This Splunk rule can track activity over time, filter out events using the conditions (such as removing known bots),
and generate meaningful alerts, making it far more effective in identifying attack behaviors.
We commit to generating such SIEM-specific rules using our \name. Among the SIEM vendors, we target Splunk vendors as Splunk~\cite{splunk, Splunk-rules} is widely used by organizations in practice in literature and our professional experience.  
Additionally, we consider the problem of rule conversion when Splunk SIEM sometimes should be migrated into another SIEM like Microsoft Sentinel~\cite{microsoftMicrosoftSentinel}.

\subsection{Motivation Scenarios}

\begin{figure}[htbp]
\setlength{\abovecaptionskip}{0pt}
\setlength{\belowcaptionskip}{0pt}
    \centering
    \scalebox{0.7}{
\includegraphics[width=0.99\linewidth]{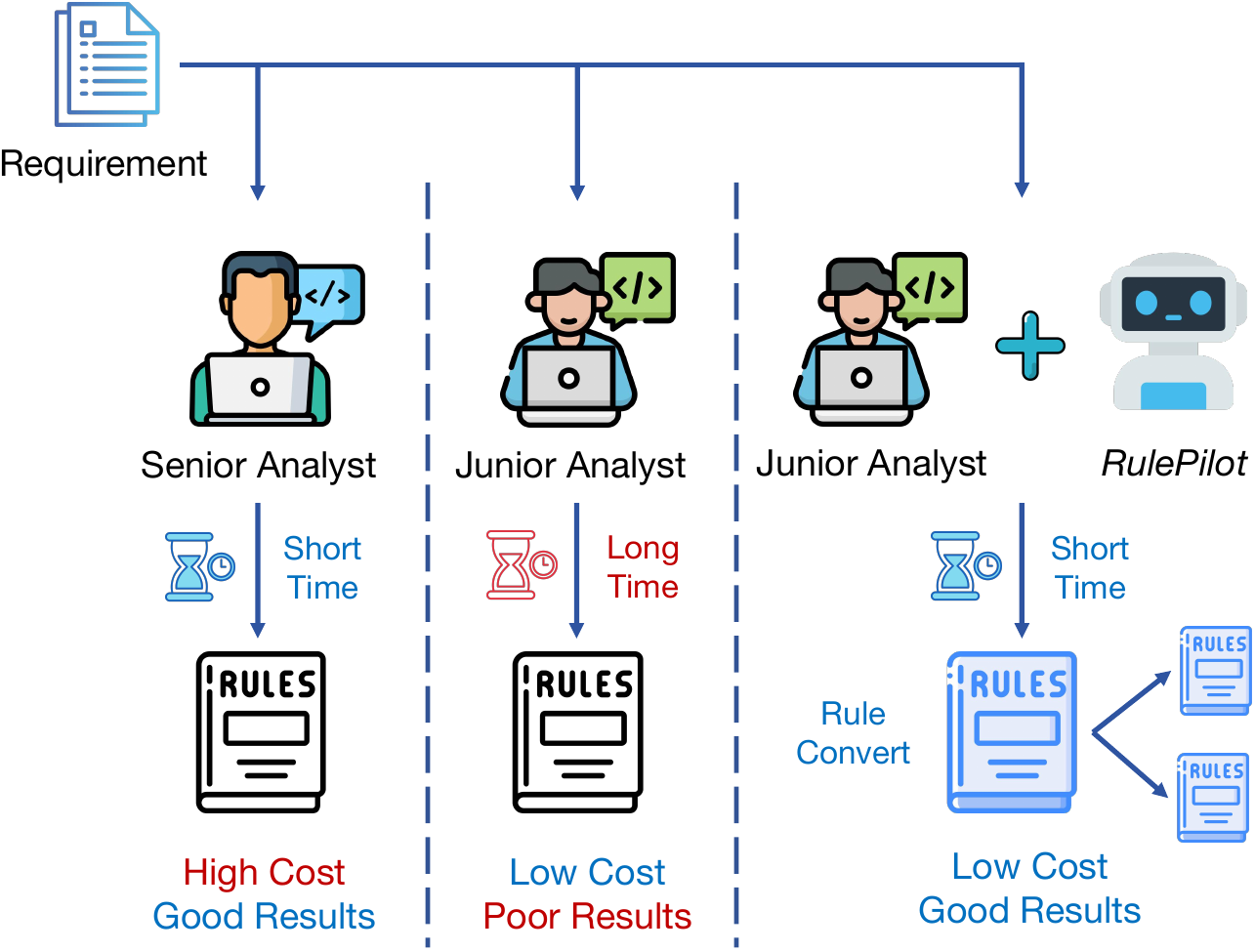}}
    \caption{Motivation scenario~\xm{\cite{SOC-motivating-example, kersten2025test:motivating-example}}: \xm{By combining with \name, junior analysts can generate concise detection rules and conversions, significantly reducing the workload of senior experts.}}
    \label{fig: motivation}
\end{figure}

As shown in Figure~\ref{fig: motivation}, existing methods require analysts to write rules manually with an attack description. There are two types of analysts: senior analysts and junior analysts. A senior analyst has rich experience and years of writing rules. They can complete the task in a short time, and the rules are of good quality. However, the cost is very high due to training and salaries. A junior analyst may lack experience. They take a long time to write rules, and the results may not be good. This motivates the use of a \name based on LLMs. \name assists in writing rules, saving time, reducing costs, and achieving better results, only with the help of junior analysts for somewhat condition handling.    
Companies often encounter system migration challenges, such as adapting validated rules to a new SIEM platform. \name addresses this by offering a rule conversion function, enabling seamless rule adaptation across different SIEM systems (e.g., from Splunk to Microsoft Sentinel). This ensures that rules generated by \name remain reusable, minimizing manual effort and streamlining future migrations.
\xm{Note that \name is designed to generate rules autonomously without human intervention. We acknowledge that human verification remains essential for final deployment. In practice, human operators validate the generated rules to ensure their correctness and effectiveness. 
The junior operator here is expected to be familiar with the SIEM environments. Compared to manual rule creation, the human role here focuses on validation, eliminating the need to master complex rule grammars.}

\subsection{LLM-Based Agents} 
An agent can be broadly defined as an autonomous entity that perceives its environment, makes decisions based on its goals, and takes actions to affect its surroundings\cite{russell2016artificial}. 
These features are inspired by human cognition and allow agents to behave consistently and effectively in dynamic, complex environments~\cite{wang2024survey}.
Researchers have been exploring machine learning techniques to automate aspects of rule generation in cybersecurity. For example, Raff et al.~\cite{raff2020automatic} introduced AutoYara, a tool that utilizes biclustering algorithms to automatically generate YARA rules for malware detection. In another study, Saxe~\cite{Saxe2020} developed YaraML, a machine learning-based toolkit designed to automate the creation of YARA rules. However, they always focus on generic YARA rules, while ignoring the customized SIEM-specific rules.
Applying LLM-based agents to generate complex detection rules, such as Splunk-specific rules, faces several significant challenges: 1) deep domain knowledge requirement. This involves meticulously analyzing rule structures step by step and crafting modular designs to guide the LLM in making precise plans and reasoned decisions.  and 2) workflow design and live SIEM integration. Developing an agent workflow capable of handling multi-step reasoning and integrating the SIEM systems is equally demanding.

%% file: 2-approach.tex
\begin{figure*}[htbp]
\setlength{\abovecaptionskip}{0pt}
\setlength{\belowcaptionskip}{0pt}
    \centering
    \scalebox{0.99}{
\includegraphics[width=0.99\linewidth]{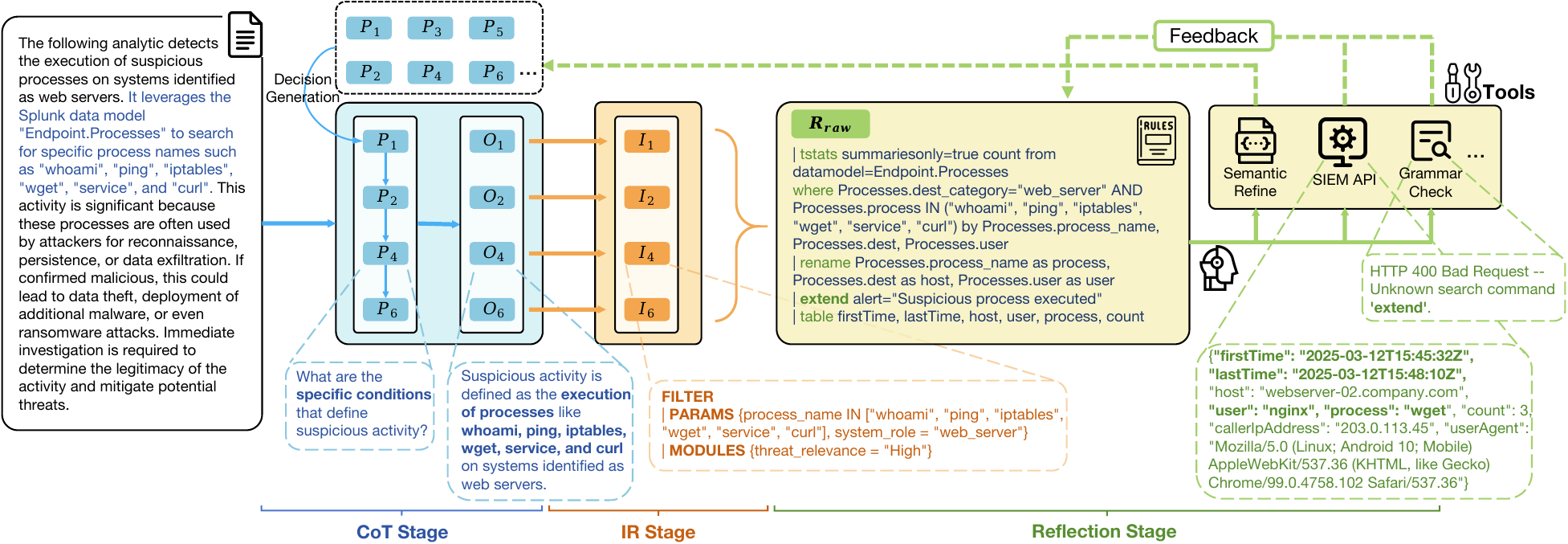}}
    \caption{Overview of \name, where \name incorporates the Chain-of-Thought (CoT), Intermediate Representation (IR), and Reflection components. \wht{Given the NLP-based input requirements, \name generates SIEM-specific rules that can be directly executed in the SIEM to detect malicious logs and trigger alerts (i.e., the malicious log).}}
    \label{fig:overview}
\end{figure*}


\section{\name: Methodology} 


\subsection{Workflow Design} \label{sec:design}

\noindent\textbf{Overview.}
As shown in Figure~\ref{fig:overview}, \name consists of three key components: Chain of Thought (CoT) reasoning, Intermediate Representation (IR), and Reflection \& Iterative Optimization. 
Given an initial rule generation request $R$, \name applies Least-to-Most Prompting (LMP) \cite{zhou2023leasttomostpromptingenablescomplex} to incrementally decompose rule generation into a structured sequence of reasoning steps $P$. Each step produces an intermediate representation that encodes its core logic, and once all steps are processed, these representations are aggregated to form an initial rule $R_{raw}$. The rule then undergoes reflection and iterative optimization, where weaknesses are identified and refined through reasoning adjustments, syntax validation, and execution-based feedback. This structured analyze-generate-reflect-optimize cycle ensures that the final rule $R_{final}$ is both logically sound and execution-efficient. 

\noindent\textbf{Chain of Thought Reasoning.}
The predefined CoT reasoning steps that \name can select and execute are below:

\noindent\textit{\textcircled{1} Interpreting the security objectives, \textcircled{2} Identifying data/log sources, \textcircled{3} Defining initial filters/conditions, \textcircled{4} Extracting relevant fields, \textcircled{5} Performing data aggregation, \textcircled{6} Optimizing the rule}

Given a rule generation request $R$, \name first decomposes it into a sequence of structured reasoning steps $P = \{p_1, p_2, ..., p_n\}$. The decomposition follows these two stages: (1) Step Selection: The model selects relevant steps from the set of predefined steps above, which cover the comprehensive aspects of rule generation.   
(2) Stepwise Execution: The model answers each subproblem in sequence, using the output of $p_i$ as contextual input for $p_{i+1}$, ensuring a gradual and structured rule refinement process.


\noindent\textbf{Intermediate Representation.}
A well-designed IR abstraction provides a clear and structured way to express rule intention.
SIEM-specific rule logic is inherently complex, requiring structured filtering, aggregation, and anomaly detection across various log sources. 
The syntax of SIEM rule languages, such as Splunk Processing Language (SPL) and Microsoft Sentinel Kusto Query Language (KQL), is highly complex, making direct generation challenging. By incorporating an intermediate representation, \name enables the model to prioritize semantic logic over syntactic details, streamlining the reasoning process and improving rule generation accuracy.

Specifically, each reasoning step corresponds to one or more IR statements, with each IR statement representing a single processing unit (pipe) in the rule. 
Given an input rule request description, the CoT process generates a sequence of reasoning steps $P = \{p_1, p_2, ..., p_n\}$, where each step $p_i$ maps to an IR component $I_i$. Formally, this can be represented as:
\begin{equation}\nonumber
\footnotesize
    I = \{I_1, I_2,...,I_n\}=\bigcup_{i=1}^{n} \mathcal{T}(o_{p_i})
\end{equation}

\noindent where $\mathcal{T}$ is the transformation function that maps each intermediate analysis output $o_{p_i}$ to an IR $I_i$.
The final IR set $I$ is the union of all individual IR components generated through this mapping process.

Once the full set of IR components $I$ has been constructed,
the initial raw rule $R_{raw}$ is generated by integrating both the semantic insights from $o_{p_i}$ and the structured transformations from $I_i$. 
\begin{equation}\nonumber
    \footnotesize
    R_{raw}=\mathcal{C}(O,I)
\end{equation}

\noindent where $\mathcal{C}$ is the rule construction function that synthesizes the intermediate analyses $O$ and the IR $I$ into an executable rule.

\noindent\textbf{Reflection and Iterative Optimization.}The optimization process incorporates a dynamic iterative mechanism, starting with an automated reflection function $\Phi(R_{raw})$, which analyzes the generated rule to identify logical inconsistencies, knowledge gaps, or syntax errors. After each iteration, the \name autonomously selects and invokes appropriate tools to address deficiencies. If the system detects unresolved issues, it triggers another refinement cycle, iterating until the rule is logically coherent, semantically accurate, and syntactically valid.
\begin{equation}\nonumber
\footnotesize
    S = \{s_{lc},s_{sc},s_{ev}\} = \Phi(R_{raw})
\end{equation}
where $S$ represents the set of identified issues:

\begin{itemize}[fullwidth,itemindent=0em]
    \item Logical Consistency ($s_{lc}$): Verifies whether all filtering, aggregation, and correlation steps align with the intended detection logic.
    \item Syntax Correctness ($s_{sc}$): Confirms that the rule adheres to the syntax requirements of the target SIEM system, such as Splunk SPL or Microsoft KQL.
    \item Execution Validity ($s_{ev}$): Ensures that the rule is structured for efficient query execution without excessive computational overhead.
\end{itemize}

\subsection{Detailed Construction}
Here, we show how to tailor our workflow to the Splunk SIEM, generating SPL rules.  

\noindent\textbf{Chain of Thought Reasoning.}
\wht{We keep the core CoT workflow unchanged, and incorporate} 
additional Splunk-adapted design elements to improve log source selection, syntax correctness, and execution efficiency. A detailed example of the CoT prompt structure is shown in Table~\ref{tab:prompt-template}. We open-source the prompts for LLMs corresponding to each function of these processes on the website~\cite{googleRulePilotDataset}.
\xm{Our prompt strategy is motivated by empirical tuning and expert insights, effectively mimicking expert-level expertise into the agent’s behavior. We used components of Identity, Instructions, Examples, and Context because they reflect how SIEM experts retrieve historical cases during manual rule construction. We also guide with DOs and DON’Ts, helping the model understand both what to do and what to avoid, based on OpenAI’s official guide~\footnote{https://platform.openai.com/docs/guides/text?api-mode=chat}.}

\begin{table}[h!]
\setlength{\abovecaptionskip}{0pt}
\setlength{\belowcaptionskip}{0pt}
\caption{Structure of the Prompt Template\label{tab:prompt-template}}  
\centering
\begin{tabularx}{\linewidth}{X} 
    \toprule[\thickline]
    \textbf{CoT Prompt Template}\\
    \midrule[\thinline]
    You are a security analyst at a cybersecurity company, specializing in writing and optimizing Splunk rules (SPL) for threat detection. \\
    Task: \textless Iterate through the tasks in the Task List\textgreater \\
    Instruction: \textless Specific Guidance such as possible keywords\textgreater \\
    Example Input: \textless Provide an example rule description\textgreater \\
    Example Output: \textless Corresponding SPL detection rule\textgreater \\
    \midrule[\thinline]
    \textbf{Task List}\\
    \midrule[\thinline]
    \textcircled{1} Map security objectives to Splunk event sources\\
    \textcircled{2} Determine necessary log fields\\
    \textcircled{3} Define efficient filtering conditions\\
    \textcircled{4} Apply field extractions and transformations\\
    \textcircled{5} Perform aggregations and anomaly detection \\
    \textcircled{6} Optimize query execution and validate syntax\\
    \bottomrule[\thickline] 
\end{tabularx}
\end{table}


\noindent\textbf{Intermediate Representation.}
We create the IR structure tailored to Splunk SPL, whose structure follows a three-part format below.   

\begin{equation}\label{equ:dsl-format}\nonumber
\footnotesize
    ⟨KEYWORD⟩ | PARAMS\{k_i = v_i\} | MODULES\{m_j\}
\end{equation}

\noindent where:
\begin{itemize}[fullwidth,itemindent=0em]
    \item $KEYWORD$ represents the core function of each rule step, including filtering logs, extracting fields, performing aggregations, or applying transformations. These IR keywords are summarized based on an extensive analysis of open-source Splunk rule sets and proprietary rules from industry collaborators. Each $KEYWORD$ corresponds to one or more SPL commands.
    Table~\ref{tab:dsl-keyword} presents the distribution and frequency of the 15 predefined IR keywords in SPL, along with their associated SPL commands.    
    \item $PARAMS$ serves as the core configuration of the rule, defining mandatory elements such as log sources, filtering conditions, and time constraints. These parameters ensure that the rule is executed within the correct context. For example, specifying \textit{index="auth\_logs" source="WinEventLog:Security"} ensures that the rule retrieves logs from relevant data sources, preventing inefficiencies caused by querying unrelated logs. Similarly, including a time constraint like \textit{earliest=-15m latest=now} helps narrow the search scope, significantly improving query speed. 
    \item $MODULES$ introduces functional annotations that enrich the interpretation of a rule, improving its flexibility, readability, and adaptability during the transformation into an executable query. Unlike $PARAMS$, which strictly define the necessary elements for rule execution, $MODULES$ describe the intended logic and analytical operations that should be applied to the retrieved data.
    For example, a module may specify "track user behavior across sessions" or "identify repeated failed login attempts", helping to capture the intent behind the rule rather than just its execution details.
\end{itemize}

Table~\ref{tab:dsl-example} shows an example on how our IR corresponds to an executable Splunk query, where the IR abstraction defines the detection logic in a structured and interpretable way, with its SPL counterpart represents the actual execution in Splunk.    
The $FILTER$ specifies where to retrieve logs, $PARAMS$ ensures correct data scoping and $MODULES$ encapsulates the detection intent, guiding how the rule should process events.

\begin{table}[h!]
\setlength{\abovecaptionskip}{0pt}
\setlength{\belowcaptionskip}{0pt}
\caption{Example for IR Statement to a Splunk Rule Pipe\label{tab:dsl-example}}  
\centering

\begin{tabularx}{\linewidth}{X} 
    \toprule[\thickline]
    \textbf{IR Example}\\
    \midrule[\thinline]
    FILTER \\ 
    \textbar ~PARAMS\\
    \{index="auth\_logs",~source="WinEventLog:Security", earliest=-30m\} \\
    \textbar~ MODULES\\
    \{"Aggregate~login~attempts", "Detect~brute~force~login~attempts"\}\\
    \midrule[\thinline]
    \textbf{Corresponding Splunk SPL Pipe}\\
    \midrule[\thinline]
    index="auth\_logs" source="WinEventLog:Security" earliest=-30m \\ 
\textbar~stats count by src\_ip \\
\textbar~where count \textgreater 10 \\
    \bottomrule[\thickline] 
\end{tabularx}
\end{table}

\begin{table}[]
\setlength{\abovecaptionskip}{0pt}
\setlength{\belowcaptionskip}{0pt} 
\renewcommand\tabcolsep{5.9pt}
\footnotesize
\caption{IR keywords along with their SPL Commands.} 
\label{tab:dsl-keyword}
\begin{tabular}{llp{3.8cm}lc}
\toprule
\textbf{Keyword}   &  & \textbf{SPL Command}                                                                           &  & \textbf{Frequency} \\ \cmidrule{1-1} \cmidrule{3-3} \cmidrule{5-5} 
FILTER    &  & search, where, eval, match                                     &  & 3129      \\
EXTRACT   &  & rex, spath, extract, kv                      &  & 2063      \\
AGGREGATE &  & stats, timechart, eventstats, tstats              &  & 1872      \\
OUTPUT    &  & table, fields,outputlookup, return                                     &  & 1609      \\
TRANSFORM &  & eval, replace, convert, fillnull &  & 1015      \\
RENAME    &  & rename                     &  & 802       \\
LOOKUP    &  & lookup,inputlookup,outputlookup                       &  & 315       \\
BUCKET    &  & bin, bucket                           &  & 91        \\
JOIN      &  & join, appendcols, transaction                &  & 83        \\
FILL      &  & fillnull, coalesce, replace                                &  & 75        \\
APPEND    &  & append, union, appendpipe                             &  & 46        \\
SORT      &  & sort, reverse                    &  & 38        \\
DEDUP     &  & dedup, uniq                                  &  & 28        \\
APPLY     &  & apply, fit                        &  & 24        \\
DEBUG     &  & noop, logtrace, dump, sendemail                           &  & 17        \\ \bottomrule
\end{tabular}
\end{table}

\xm{To address SIEM-specific edge cases, we design the IR with explicit support. 
For nested operators, which are commonly used in filter and transformation logic, we incorporate two strategies: For simple constructs (e.g., eval, match, where), we allow controlled nesting within a single IR statement; for more complex expressions involving multi-layer joins or condition chaining, we enforce decomposition into multiple atomic IR statements to preserve interpretability and reduce error propagation during transformation.}
 

\xm{To accommodate SIEM-specific syntax variations of multiple SIEMs, our IR incorporates a pluggable keyword dictionary architecture. For each target SIEM system (e.g., Splunk SPL, Microsoft KQL), we can curate and maintain a dedicated dictionary of IR keywords and associated translation templates derived from empirical rule corpora and vendor documentation. This allows the IR-to-query compiler to flexibly adapt the output semantics and syntactic form of different SIEMs. 
}

\noindent\textbf{Reflection and Iterative Optimization.} 
\xm{Different from LLM-based self-debugging~\cite{chen2023teachinglargelanguagemodels:self-debugging, tian2024debugbenchevaluatingdebuggingcapability-debugging} that typically relies on prompting-based re-generation conditioned on observed errors or exceptions, our reflection adopts a scoring-based multi-layered mechanism. This enables the system to iteratively reconstruct faulty rule components at both the IR and final SPL levels, addressing deeper issues such as semantic gaps, abstraction mismatches, logical inconsistencies, field coverage, and execution viability, rather than merely rewriting surface text. Our reflection mechanism integrates semantic-level diagnostics, real execution feedback via selective modules, and a scoring-based evaluation to further analyze the  logical consistency, field coverage, and execution viability using a scoring-based evaluation ($s_{lc}$, $s_{sc}$, $s_{ev}$), and selectively invokes CoT-based refinement and SIEM-integrated validation routines.}


\wht{Specifically, \name first performs syntax validation using Splunklib’s dry-run mode~\cite{githubSplunksdkpythonsplunklibMaster}, which allows the system to check for syntax errors without executing the query. Second,} \name dynamically invokes a set of predefined optimization modules to address identified logical/structural inconsistencies. If $\Phi(R_{raw})$ detects any of $\{s_{lc}, s_{sc}, s_{ev}\}$, indicating logical inconsistencies, structural violations, or execution failures, the system applies targeted refinements through two steps:
\begin{enumerate}[fullwidth,itemindent=0em]
    \item CoT-Based Refinement Modules ($M_{CoT}$): The system revisits earlier CoT reasoning steps and regenerates the affected IR statements, refining the rule logic and improving structural coherence. This produces an updated intermediate rule $R'_{raw}$, defined as:
        \begin{equation}\nonumber
        \footnotesize
            R'_{raw}=M_{CoT}(R_{raw}, s_{lc})
        \end{equation}
        \item \wht{Rule Refinement Modules} ($M_{Splunk}$): \name integrates with live Splunk’s validation and execution environment. The rule is executed via Splunk’s API, retrieving real log data. If the results deviate from the intended detection objective, \name iteratively refines the SPL by adjusting filters, modifying conditions, or restructuring query logic based on execution feedback.  
        \begin{equation}\nonumber
        \footnotesize
            R_{final} = M_{Splunk}(R'_{raw}, S)
        \end{equation}
\end{enumerate}


\subsection{Rule Conversion}

To ensure the interoperability and adaptability of \name across different SIEM platforms, we implement a Rule Conversion Module that translates rules from one SIEM vendor (e.g., Splunk SPL) into another (e.g., Microsoft KQL). This conversion process is designed to preserve the logical intent of the original rule while adapting it to the syntax, function names, and query structures of the target SIEM.

As shown in Algorithm~\ref{alg:rule_conversion}, the Rule Conversion follows a structured multi-step approach, 
beginning by segmenting the input rule into individual pipes, and then breaking down a complex task into smaller, more manageable units.
However, this breakdown may cause a loss of contextual dependencies between pipes, so an LLM-based function extraction module is introduced to retrieve each pipe’s purpose and variable mappings, ensuring coherence in later stages.   
Once the semantic information is extracted, the system converts each pipe sequentially, allowing previously processed pipes to provide contextual support.

To further improve accuracy, \xm{we incorporate a Retrieval-Augmented Generation (RAG)~\cite{DBLP:conf/nips/LewisPPPKGKLYR020:RAG-first} mechanism that dynamically maps keywords and functions from the source SIEM to their equivalents in the target SIEM.}
\xm{The RAG knowledge base is bootstrapped from Microsoft’s official migration documentation \cite{microsoftMigrateSplunk}, which provides detailed mappings between Splunk detection rules and their KQL counterparts. From this corpus, we extract every \texttt{<SPL command, KQL operator, usage snippet>} triple, normalise aliases (e.g., \texttt{rex} $\leftrightarrow$ \texttt{extract}) and build a key-value mapping database that aligns functionally equivalent operations across the two SIEMs. Each SPL command string and its descriptive context are embedded with the \texttt{text-embedding-ada-002}~\cite{openAI-vector-model}.}
\xm{During conversion, every pipe is first tokenised into \textit{<verb, args, fields>} tuples.
The verb plus surrounding comments are embedded on-the-fly, and a top-$k$ (default $k=10$) vector search is issued. Candidates with cosine similarity above a threshold (i.e., 0.82 used) are retained and re-ranked with a BM25~\cite{bm25} lexical score to favour exact-string matches.
If the SPL command has a high-confidence match, the retrieved KQL operator (and an example usage) is attached to the LLM prompt as a structured ``conversion hint''.}
This retrieval-augmented approach ensures that the LLM does not solely rely on pre-trained knowledge but is instead guided by vendor-specific best practices and real-world rule patterns.

\begin{algorithm}
\footnotesize
\caption{Rule Conversion from SIEM Vendor A to B}
\label{alg:rule_conversion}
\begin{algorithmic}[1]
\REQUIRE Rule $R_A$ from SIEM Vendor A, Target SIEM Vendor B
\ENSURE Converted rule $R_B$ for SIEM Vendor B

\STATE \textbf{Step 1: Pipe Segmentation}
\STATE Split $R_A$ into a sequence of pipes: $P = \{p_1, p_2, ..., p_n\}$

\STATE \textbf{Step 2: Function Extraction}
\FOR{each pipe $p_i \in P$}
    \STATE $(f_i, in_i, out_i) \gets \text{ExtractFunctionInfo}(p_i)$
\ENDFOR

\STATE \textbf{Step 3: Context-Aware Pipe Conversion}
\FOR{each pipe $p_i \in P$}
    \STATE $P_{\text{prior}} \gets \text{GetPriorPipes}(p_i, P)$
    \STATE $K_i \gets \text{RetrieveKeyword}(p_i, \text{Vendor A}, \text{Vendor B})$
    \STATE $p'_i \gets \text{ConvertPipe}(p_i, in_i, out_i, P_{\text{prior}}, K_i)$
    \STATE $P' \gets P' \cup \{p'_i\}$
\ENDFOR

\STATE \textbf{Step 4: Assemble Converted Rule}
\STATE $R_B \gets \text{AssembleRule}(P')$

\RETURN $R_B$
\end{algorithmic}
\end{algorithm}

%% file: 3-evaluation.tex
\section{Evaluation}

In this section, we aim to evaluate the following research questions.

\begin{itemize}[fullwidth,itemindent=0em]
    \item RQ1-\textbf{Accuracy}: 
    \wht{How effective is \name in generating SIEM-specific detection rules, measured by similarity to official rules and execution success across SIEM vendors?}

    \item RQ2-\textbf{Efficiency}: What are the latency and resource costs of the rule generation process?   
    \item RQ3-\textbf{Ablation Study}: Do the specific
    components in \name help improve the quality of rule generation?   
    \item RQ4-\textbf{Compatibility}: Does \name support the conversion between Splunk SPL and Microsoft KQL? 
\end{itemize}

\subsection{Experimental Settings}\label{sec:settings}

\subsubsection{Implementation Details} 
We implemented a fully functional prototype of \name, designed for automated rule generation and conversion in Splunk SIEM. 
\name is built upon GPT-4o, DeepSeek-V3 (671B), and LLaMA-3 (405B), with agent-based orchestration implemented using their function-calling capabilities. To control generation behavior, we configure all models with a temperature of 0.3 (balancing determinism and flexibility), top-p of 0.9 (ensuring controlled diversity), and set a maximum response length of 512 tokens to prevent excessively long outputs.
We use the Splunk of version 9.3.1, and the trial license for experiments.

\subsubsection{Datasets}
We evaluate \name using two sources of rules: \textbf{Splunk official rules} from the Splunk Security Content repository~\footnote{\url{https://github.com/splunk/security_content}. We primarily use rules from the \textit{detections} folder to evaluate the similarity score and reference macro definitions from the \textit{macros} folder for completeness.} as the ground truth to evaluate the similarity score, and the \textbf{custom rules between Splunk SPL and Microsoft KQL currently used by our industry collaborator} for their security operations to evaluate the compatibility.  
The Splunk official rules we download contain a total of 1,699 samples, organized into five major categories based on their focus (shown in Table~\ref{tab:datasets}),  \xm{maximizing coverage across comprehensive and diverse categories, reducing evaluation bias.} 
The key components of the datasets include the rule body (SPL) and the corresponding description that explains the purpose and context of the rule. 
The original datasets contained macros, which are vendor-specific or environment-specific variations. The macro abstracts the specific directives, and do not conform to rule grammars, possibly affecting rule consistency.
To ensure a fair comparison, we replaced all macros with standardized definitions based on Splunk’s official macro library. 
For example, a macro like \colorbox{mygray}{\texttt{'process\_cmd'}}, should be replaced with its specific SPL equivalent: \texttt{Processes.process\_name = cmd.exe}. 

\subsubsection{System Log Collection} 
To evaluate the execution success of our generated rules, we simulate various atomic attacks~\cite{atomic-test-mitre} provided by MITRE ATT\&CK, including 229,968 system logs from 61 atomic tests, covering 12 tactics in MITRE ATT\&CK, \xm{covering broad applicability and reducing evaluation bias.} 
The system logs were collected using EventViewer in a controlled environment, capturing system, Sysmon, and PowerShell logs on a virtual machine running Windows 10 (64-bit).
Our evaluation focuses primarily on Windows events due to their widespread use in both enterprises and consumer marketss~\cite{windows-more-target}.
The datasets have their labels based on our simulation process, open-sourced in~\cite{googleRulePilotDataset}.

\begin{table}[htbp]
\centering
\setlength{\abovecaptionskip}{0pt}
\setlength{\belowcaptionskip}{0pt}
\caption{Category of our ground-truths, with each item containing the NLP descriptions and associate SPL rules.}  
\label{tab:datasets}
\renewcommand\tabcolsep{15.2pt} 
\footnotesize 
\begin{tabular}{llll}
\toprule
\multicolumn{1}{c}{Rules-Set Type} &  & \multicolumn{1}{c}{size} & \multicolumn{1}{c}{Time Frame}  \\ \cmidrule{1-1} \cmidrule{3-4} 
Application                        &  & 125                      & 2024-09-30 -- 2024-11-19                   \\ \cmidrule{1-1} \cmidrule{3-4} 
Cloud                              &  & 271                      & 2024-09-30 -- 2024-10-31                   \\ \cmidrule{1-1} \cmidrule{3-4} 
Endpoint                           &  & 1187                     & 2024-09-24 -- 2024-12-03                \\ \cmidrule{1-1} \cmidrule{3-4} 
Network                            &  & 44                       & 2024-09-25 -- 2024-11-06                   \\ \cmidrule{1-1} \cmidrule{3-4} 
Web                                &  & 72                       & 2024-09-30 -- 2024-10-17                  \\ \bottomrule
\end{tabular}
\end{table}



\subsubsection{Baseline} 
We compare \name's performance against that based upon the standalone LLMs of GPT-4o, DeepSeek-V3 (671B), and LLaMa-3 (405B), without the structured reasoning and function-calling mechanisms of \name.
For a fair comparison, both baseline models generate rules using the same prompts as those employed by \name in its rule generation step, without any additional multi-step processing, validation, or refinement. 
We download DeepSeek-V3 and LLaMa-3 models from Hugging Face~\cite{huggingfaceDeepseekaiDeepSeekV3Hugging},\cite{huggingfaceMetallamaLlama31405BHugging}
To ensure consistency across all models, we use the same model parameters as those employed by \name.


\subsection{Evaluation Metrics}
\noindent\textbf{Accuracy.} 
Our accuracy evaluation consists of two complementary metrics: \textit{quantifiable similarity assessment} and \textit{an LLM-based evaluator}. 
The first metrics measure the textual similarity between the generated rules and the ground truth (i.e., the official rules), creating an objective assessment of structural and lexical alignment. We adopt three well-established quantifiable metrics below.  
\begin{itemize}[fullwidth,itemindent=0em]
    \item \textit{ROUGE} (Recall-Oriented Understudy for Gisting Evaluation)~\cite{lin2004rouge} is an NLP metric that compares machine-generated text with reference text to measure content similarity. A higher ROUGE-k indicates a greater overlap of k-grams between the generated and ground truth rule. Here, we set k = 1 and include ROUGE-L, capturing the longest common subsequence to reflect structural alignment. 
    \item \textit{BLEU} (Bilingual Evaluation Understudy)~\cite{papineni2002bleu} is a widely used precision-oriented NLP metric that evaluates text similarity based on n-gram overlap. A higher BLEU-k score indicates better alignment between the generated and ground truth rule. We use BLEU-4 in our evaluation. 
    \item \textit{METEOR} (Metric for Evaluation of Translation with Explicit ORdering)~\cite{banerjee2005meteor} is an advanced NLP metric that improves upon BLEU by incorporating stemming, synonym matching, and word order considerations, 
    making it a more robust metric for comparing variations in rule expressions. 
\end{itemize}

The \textit{quantifiable similarity assessment} offers an objective metric but may yield misleadingly high scores for syntactically similar yet semantically incorrect rules. To mitigate this, we adopt the LLM-as-a-judge approach, a scalable and explainable method for approximating human preferences~\cite{li2024llmsasjudgescomprehensivesurveyllmbased}. 
We evaluate rule quality from a semantic perspective, considering six key evaluation dimensions below.

\begin{itemize}[fullwidth,itemindent=0em]
    \item \textit{Logical Consistency} (LC). 
    The LLM looks at conditions, operators, and filters to see if anything is missing or changed. 
    \item \textit{Syntax Correctness} (SC). 
    The LLM checks for mistakes in the query and looks for ways to write it better.
    \item \textit{Readability \& Maintainability} (RM). 
    The LLM checks if the rule is written in a clear way, without unnecessary complexity. 
    \item \textit{Condition Coverage} (CC). 
    The LLM ensures no key conditions are missing or unnecessary constraints are added.
    \item \textit{False Positive \& False Negative Risk} (FPFNR). 
    The LLM checks if the rule is too strict (which may miss real threats) or too loose (which may flag normal activities).
    \item \textit{Execution Efficiency} (EE). 
    The LLM analyzes whether the rule uses complex operations, unnecessary filters, or inefficient queries that could slow down processing. 
\end{itemize}

\begin{figure}[htbp]
\setlength{\abovecaptionskip}{0pt}
\setlength{\belowcaptionskip}{0pt}
    \centering
\includegraphics[width=0.30\linewidth]{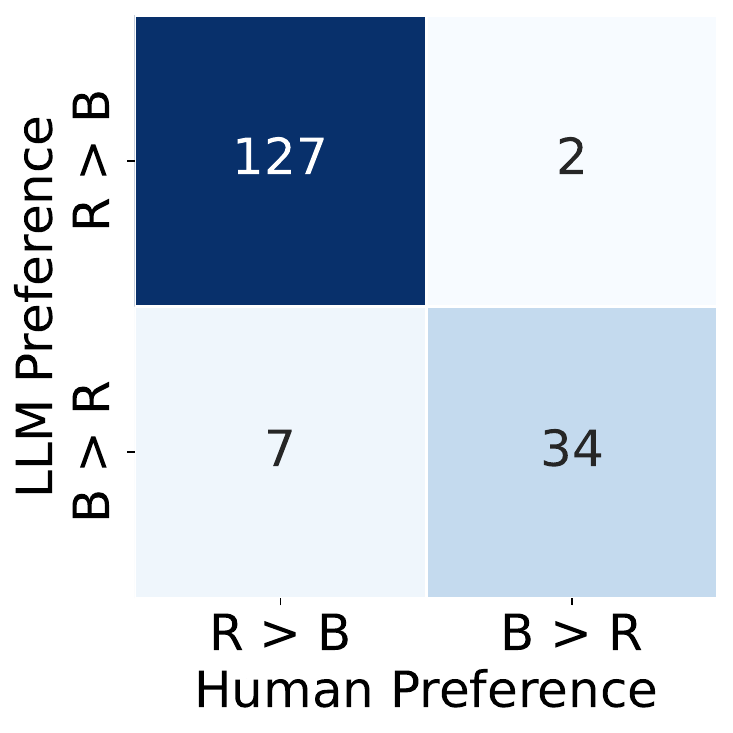}
    \caption{\xm{Preference inter-rater alignment between the LLM and the human evaluator. \textbf{R} refers to \name, and \textbf{B} refers to \textbf{Baseline}. The “$>$” indicates that the method was rated higher in a pairwise comparison. The diagonal entries indicate agreement between the LLM and the human evaluator.}}
    \label{fig:confusion_matrix}
\end{figure}

\noindent\wht{
We adopt a scoring scheme ranging from 0 to 1 for each evaluation dimension. Instead of focusing on absolute scores, we emphasize relative rankings across outputs under the same prompt for evaluating the effectiveness of different methods.
To mitigate evaluation bias, we followed a human-aligned iterative evaluation framework. An experienced human expert and an LLM were involved in an iterative prompt refinement process to align the evaluation standards.} 
\xm{We define inter-rater agreement as a match in relative preference, for example, both the human and the LLM giving higher scores to \name over the corresponding vanilla LLMs (baseline) is considered consistent, regardless of exact numerical values.
Under this definition, we show the matrix in Figure~\ref{fig:confusion_matrix} based on pairwise preferences, where our inter-rater agreement test reaches a larger Cohen’s Kappa~\cite{Cohen-Kappa} score of 0.85, indicating strong agreement. 
}

\noindent\textbf{Execution Success.} We ingest all collected logs into Splunk, execute the generated rules as search queries, and verify their accuracy by comparing retrieved logs against the ground truth from simulated atomic tests. Execution is considered successful if the retrieved logs match the expected attack-generated logs. We compute precision, recall, and other metrics to assess rule effectiveness. 
To quantify performance, we use $precision= \frac{TP}{TP+FP}$, $recall= \frac{TP}{TP+FN}$, and $F1=2 \cdot \frac{\text{Precision} \cdot \text{Recall}}{\text{Precision} + \text{Recall}}$ as evaluation metrics.


\begin{table*}[htbp]
\centering
\setlength{\abovecaptionskip}{0pt}
\setlength{\belowcaptionskip}{0pt}
\caption{\wht{Syntax-level evaluation.} Similarity comparison of the generated rules from \name (RP) and baselines (BL) with ground truth, including GPT-4o (GPT), DeepSeek-V3-671B (DS), and LLaMa-3-405B (LLaMa).} 
\label{tab:results}
\renewcommand\tabcolsep{3.4pt}
\footnotesize
\begin{tabular}{lc>{\columncolor{mygray}}cc>{\columncolor{mygray}}cc>{\columncolor{mygray}}cc|>{\columncolor{mygray}}cc>{\columncolor{mygray}}cc>{\columncolor{mygray}}cc|>{\columncolor{mygray}}cc>{\columncolor{mygray}}cc>{\columncolor{mygray}}cc|>{\columncolor{mygray}}cc>{\columncolor{mygray}}cc>{\columncolor{mygray}}cc}
\toprule
\multicolumn{1}{c}{Category} &  & \multicolumn{6}{c|}{BLEU ($\uparrow$)}                                                           & \multicolumn{6}{c|}{ROUGE-1 ($\uparrow$)}                                                        & \multicolumn{6}{c|}{ROUGE-L ($\uparrow$)}                                                        & \multicolumn{6}{c}{METEOR ($\uparrow$)}                                                         \\ \cmidrule{3-26} 
\multicolumn{1}{c}{}                   &  & \multicolumn{2}{c|}{GPT}         & \multicolumn{2}{c|}{DS}          & \multicolumn{2}{c|}{LLaMa} & \multicolumn{2}{c|}{GPT}         & \multicolumn{2}{c|}{DS}          & \multicolumn{2}{c|}{LLaMa} & \multicolumn{2}{c|}{GPT}         & \multicolumn{2}{c|}{DS}          & \multicolumn{2}{c|}{LLaMa} & \multicolumn{2}{c|}{GPT}         & \multicolumn{2}{c|}{DS}          & \multicolumn{2}{c}{LLaMa} \\
\multicolumn{1}{c}{}                   &  & RP   & \multicolumn{1}{c|}{BL}   & RP   & \multicolumn{1}{c|}{BL}   & RP           & BL          & RP   & \multicolumn{1}{c|}{BL}   & RP   & \multicolumn{1}{c|}{BL}   & RP           & BL          & RP   & \multicolumn{1}{c|}{BL}   & RP   & \multicolumn{1}{c|}{BL}   & RP           & BL          & RP   & \multicolumn{1}{c|}{BL}   & RP   & \multicolumn{1}{c|}{BL}   & RP          & BL          \\ \cmidrule{1-1} \cmidrule{3-26} 
application (54)                       &  &  39.1  & \multicolumn{1}{c|}{33.6} & 43.8 & \multicolumn{1}{c|}{30.2} & 32.5         & 31.1        & 49.2 & \multicolumn{1}{c|}{36.6} & 42.3 & \multicolumn{1}{c|}{33.9} & 41.6         & 36.8        & 41.7 & \multicolumn{1}{c|}{26.2} & 32.3 & \multicolumn{1}{c|}{24.4} & 33.2         & 26.5        & 41.3 & \multicolumn{1}{c|}{27.3} & 26.7 & \multicolumn{1}{c|}{19.6} & 29.5        & 27.1        \\ \cmidrule{1-1} \cmidrule{3-26} 
cloud (271)                            &  & 47.9 & \multicolumn{1}{c|}{33.9} & 40.9 & \multicolumn{1}{c|}{25.4} & 38.3         & 27.0        & 58.7 & \multicolumn{1}{c|}{44.4} & 53.1 & \multicolumn{1}{c|}{21.6} & 48.1         & 28.8        & 53.7 & \multicolumn{1}{c|}{37.5} & 52.5 & \multicolumn{1}{c|}{29.1} & 46.1         & 26.7        & 58.1 & \multicolumn{1}{c|}{43.8} & 61.4 & \multicolumn{1}{c|}{32.2} & 65.0        & 33.9        \\ \cmidrule{1-1} \cmidrule{3-26} 
endpoint (1,187)                        &  & 42.8 & \multicolumn{1}{c|}{29.5} & 34.8 & \multicolumn{1}{c|}{24.5} & 32.8         & 27.8        & 59.8 & \multicolumn{1}{c|}{36.6} & 51.0 & \multicolumn{1}{c|}{24.4} & 48.7         & 43.0        & 57.3 & \multicolumn{1}{c|}{32.5} & 42.0 & \multicolumn{1}{c|}{30.9} & 40.1         & 37.0        & 66.3 & \multicolumn{1}{c|}{37.8} & 35.9 & \multicolumn{1}{c|}{29.6} & 42.5        & 22.9        \\ \cmidrule{1-1} \cmidrule{3-26} 
network (43)                           &  & 41.9 & \multicolumn{1}{c|}{41.5} & 35.2 & \multicolumn{1}{c|}{27.9} & 45.4         & 39.0        & 60.1 & \multicolumn{1}{c|}{49.5} & 37.0 & \multicolumn{1}{c|}{27.5} & 58.8         & 37.7        & 57.1 & \multicolumn{1}{c|}{43.3} & 27.0 & \multicolumn{1}{c|}{18.4} & 58.2         & 30.0        & 59.2 & \multicolumn{1}{c|}{42.4} & 25.5 & \multicolumn{1}{c|}{24.9} & 60.7        & 34.9        \\ \cmidrule{1-1} \cmidrule{3-26} 
web (72)                               &  & 41.6 & \multicolumn{1}{c|}{34.8} & 27.5 & \multicolumn{1}{c|}{22.1} & 32.0         & 28.2        & 57.0 & \multicolumn{1}{c|}{43.3} & 39.8 & \multicolumn{1}{c|}{20.6} & 38.6         & 28.8        & 50.1 & \multicolumn{1}{c|}{37.9} & 31.9 & \multicolumn{1}{c|}{14.5} & 36.6         & 27.3        & 56.6 & \multicolumn{1}{c|}{41.2} & 28.9 & \multicolumn{1}{c|}{17.4} & 43.8        & 34.8        \\ \cmidrule{1-1} \cmidrule{3-26} 
Total (1,627)                                 &  & 43.4 & \multicolumn{1}{c|}{30.9} & 35.8 & \multicolumn{1}{c|}{24.8} & 34.0         & 28.1        & 59.1 & \multicolumn{1}{c|}{38.5} & 50.2 & \multicolumn{1}{c|}{24.2} & 48.2         & 39.7        & 55.9 & \multicolumn{1}{c|}{33.6} & 42.6 & \multicolumn{1}{c|}{29.3} & 41.2         & 34.3        & 63.5 & \multicolumn{1}{c|}{38.7} & 39.3 & \multicolumn{1}{c|}{29.0} & 46.4        & 25.7        \\ \bottomrule
\end{tabular}
\end{table*}

\subsection{Evaluation Results}
\subsubsection{RQ1-Accuracy} 
We present our \textit{quantifiable similarity assessment} in Table~\ref{tab:results}.
Overall, \name consistently outperforms all its corresponding baseline models across every category. The improvements range from 20.9\% to 107.4\%, demonstrating that \name significantly enhances both the syntactic accuracy of the generated rules.
Among the different detection categories, the cloud achieves the highest overall performance, 
suggests that cloud-based detection rules are relatively easier for \name to generate accurately, possibly due to the structured and well-documented nature of cloud security rules. In contrast, the web category exhibits the lowest performance across most metrics.
This indicates that web-related security rules tend to be more complex or diverse, making it harder for models to capture accurate patterns. Literature~\cite{236200}, \cite{298234} also support this claim that web-related security rules are complex.
When comparing different LLMs,
\name achieves its highest performance using GPT-4o, compared to  and LLaMa-3. 
This suggests that GPT-4o is better suited for structured rule-generation tasks, likely due to its improved reasoning and instruction-following capabilities.

\begin{figure}[htbp]
\setlength{\abovecaptionskip}{0pt}
\setlength{\belowcaptionskip}{0pt}
    \centering
\includegraphics[width=0.9\linewidth]{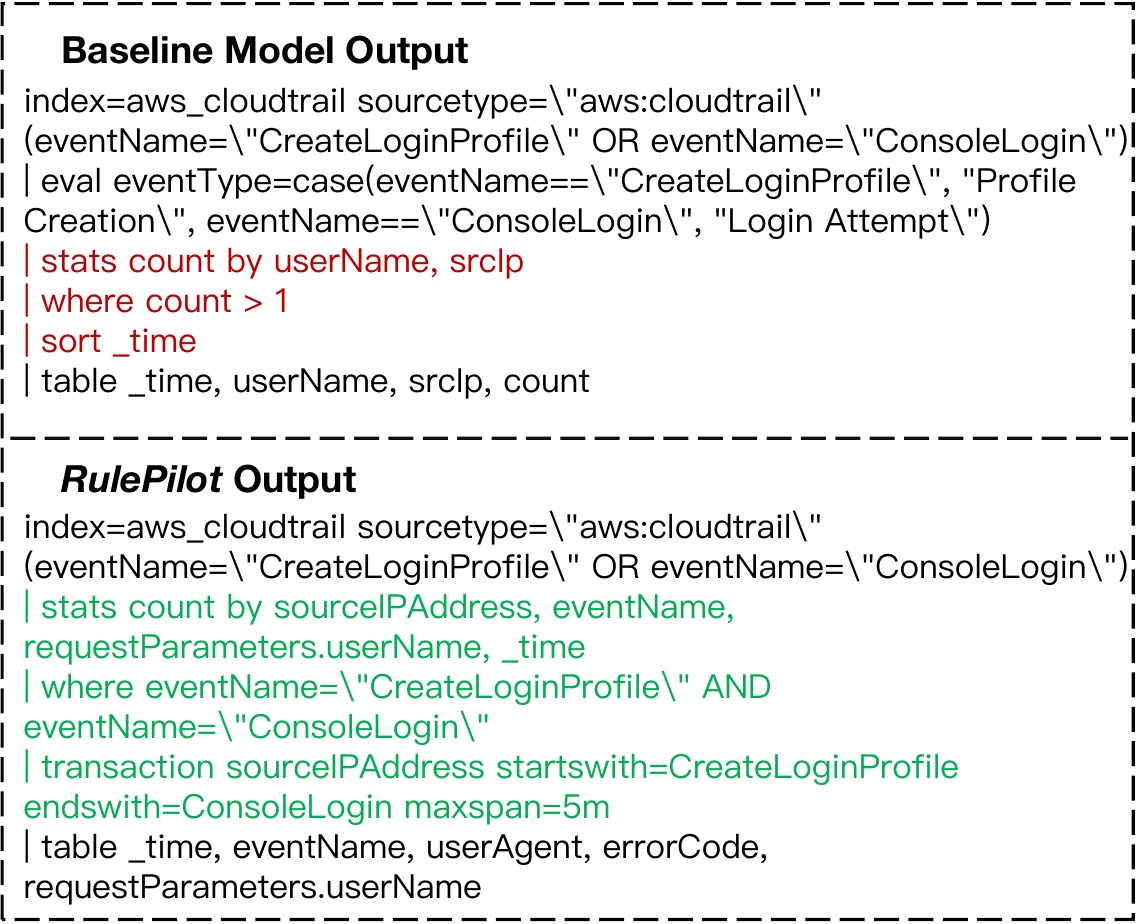}
    \caption{\wht{Expert-reviewed} comparison between rules generated by \name \wht{versus those} generated by standalone GPT-4o.} 
    \label{fig:evaluation-comparison}
\end{figure}

Looking deeply, \wht{our experts judge a concrete example of a rule generated by \name and the GPT-4o baseline model. We present an expert-reviewed result in Figure~\ref{fig:evaluation-comparison}}. 
This demonstrates a rule related to AWS CloudTrail login and profile creation monitoring. We can find that the baseline model fails to properly correlate profile creation and subsequent login attempts, relying only on counting occurrences per user and IP. While it correctly filters relevant event types, it lacks a mechanism to determine if a login actually follows a profile creation within a short window.

\begin{figure}[t]
\setlength{\abovecaptionskip}{0pt}
\setlength{\belowcaptionskip}{0pt}
\footnotesize
  \centering
    \subfigure[\texttt{GPT-4o}]{\includegraphics[width=0.15\textwidth]{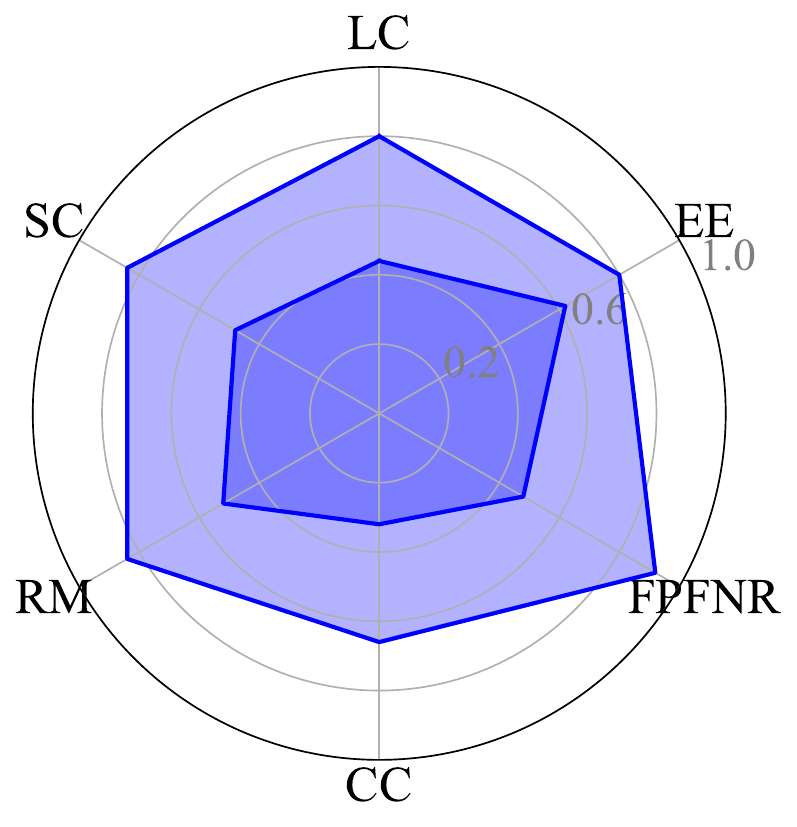}} 
    \subfigure[\texttt{DeepSeek-V3}]{\includegraphics[width=0.15\textwidth]{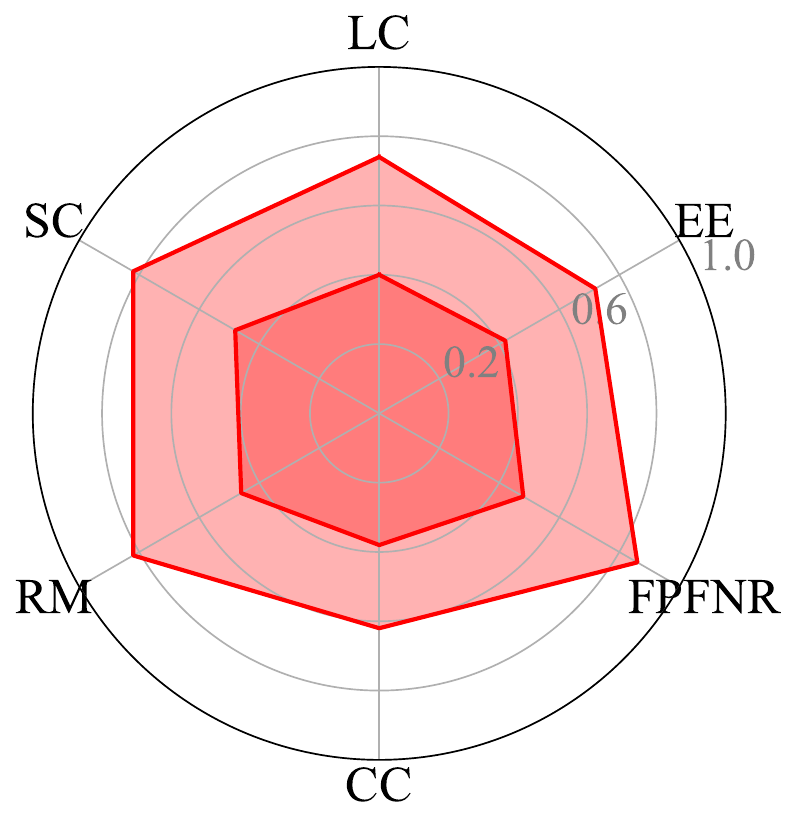}} 
    \subfigure[\texttt{LlaMa-3}]{\includegraphics[width=0.15\textwidth]{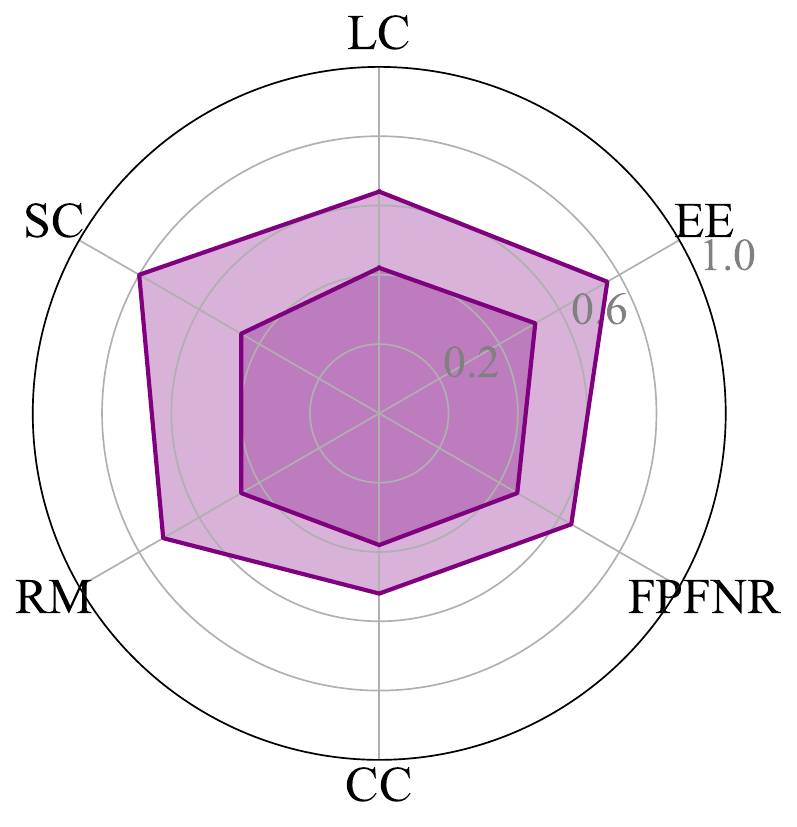}} 
\caption{\wht{Semantic-level evaluation}. Radar chart of LLM-based evaluator: the inner shaded area represents the baseline model's score, while the outer contour represents \name's score.
} 
\label{fig:radar}
\end{figure}

Second, to avoid the syntactically similar yet semantically incorrect evaluation, we show the results of radar chart in Figure~\ref{fig:radar}, illustrating the results of LLM-based evaluator results. We can find that \name consistently outperforms all standlone LLMs across six evaluation dimensions.
GPT-4o achieves the highest performance across most dimensions, particularly in Syntax Correctness (SC).
In contrast, DeepSeek-V3 and LlaMa-3 show weaker performance, especially in Condition Coverage (CC) and False Positive \& False Negative Risk (FPFNR). 

\noindent\textbf{Execution Success.} 
To show the execution success of the generated rules in a SIEM vendor, we show our results in Table~\ref{tab:redteam-result}, where we find that \name consistently achieves higher precision and recall across most tactics compared to GPT-4o, demonstrating its ability to generate executable detection rules in Splunk. Notably, \name achieves perfect (100\%) precision across multiple tactics, confirming that its generated rules accurately match ground truth detections without false positives in these cases.
However, certain tactics exhibit lower precision and recall, particularly for Privilege Escalation, where GPT-4o fails entirely (0\% precision and recall), while \name retains some detection capability. These lower scores are primarily due to the \textit{subtle} nature of malicious logs associated with these tactics, where critical identifying fields do not explicitly appear in the rule descriptions. As a result, baseline models struggle to generate effective detection rules, relying only on static descriptions without real-time feedback.

In contrast, \name demonstrates significantly better performance due to its ability to autonomously call the Splunk API, retrieving real-time log feedback and refining its rules iteratively. 
This self-reflective and API-driven approach enables \name to detect complex attack patterns that GPT-4o fails to capture, particularly in scenarios where key indicators are not directly stated in the initial rule descriptions.

\begin{table}[htbp]
\footnotesize
\setlength{\abovecaptionskip}{0pt}
\setlength{\belowcaptionskip}{0pt}
\centering
\caption{Execution-level success on the Splunk SIEM.} 
\label{tab:redteam-result}
\renewcommand\tabcolsep{5pt}
\begin{tabular}{lllllll}
\toprule[\thickline]
\multicolumn{1}{c}{Tactic} &  & \multicolumn{2}{c}{Precision (\%)} &  & \multicolumn{2}{c}{Recall (\%)} \\ \cmidrule{1-1} \cmidrule{3-4} \cmidrule{6-7} 
                           &  &  \name   & GPT-4o   &  &  \name  & GPT-4o  \\ \cmidrule{1-1} \cmidrule{3-4} \cmidrule{6-7} 
Reconnaissance             &  & \cellcolor{mygray}  1.000                       &   1.000       &  & \cellcolor{mygray}  1.000                   &  1.000       \\ \cmidrule{1-1} \cmidrule{3-4} \cmidrule{6-7} 
Initial Access &  & \cellcolor{mygray} 1.000                       &   1.000       &  &    1.000    \cellcolor{mygray}           &  1.000       \\ \cmidrule{1-1} \cmidrule{3-4} \cmidrule{6-7} 
Execution                  &  & \cellcolor{mygray}  1.000                       &   0.909       &  &  \cellcolor{mygray}  0.750                   & 0.416        \\ \cmidrule{1-1} \cmidrule{3-4} \cmidrule{6-7} 
Persistence                &  &  \cellcolor{mygray}  1.000                      &  1.000        &  & \cellcolor{mygray}  0.818                     & 0.714        \\ \cmidrule{1-1} \cmidrule{3-4} \cmidrule{6-7} 
Privilege Escalation       &  & \cellcolor{mygray} 0.600                       &  0.000        &  & \cellcolor{mygray}   0.214                    &  0.000       \\ \cmidrule{1-1} \cmidrule{3-4} \cmidrule{6-7} 
Defense Evasion            &  & \cellcolor{mygray}  0.733                       &  0.600        &  &  \cellcolor{mygray} 0.833                    & 0.656        \\ \cmidrule{1-1} \cmidrule{3-4} \cmidrule{6-7} 
Credential Access          &  & \cellcolor{mygray}  1.000                      &   1.000       &  & \cellcolor{mygray}   0.450                    &  0.264       \\ \cmidrule{1-1} \cmidrule{3-4} \cmidrule{6-7} 
Discovery                  &  & \cellcolor{mygray} 0.667                        & 0.167         &  & \cellcolor{mygray}   0.444                    & 0.100        \\ \cmidrule{1-1} \cmidrule{3-4} \cmidrule{6-7} 
Lateral Movement           &  & \cellcolor{mygray} 1.000                        & 1.000         &  & \cellcolor{mygray}  0.667                    & 0.667        \\ \cmidrule{1-1} \cmidrule{3-4} \cmidrule{6-7} 
Collection                 &  & \cellcolor{mygray} 1.000                       & 1.000         &  & \cellcolor{mygray}  1.000                     & 1.000        \\ \cmidrule{1-1} \cmidrule{3-4} \cmidrule{6-7} 
Command and Control        &  & \cellcolor{mygray} 1.000                       & 1.000         &  & \cellcolor{mygray} 1.000                     & 1.000        \\ \cmidrule{1-1} \cmidrule{3-4} \cmidrule{6-7} 
Exfiltration               &  & \cellcolor{mygray}  0.667                       & 0.333         &  & \cellcolor{mygray} 0.500                     &  0.200       \\ \cmidrule{1-1} \cmidrule{3-4} \cmidrule{6-7} 
Impact                     &  & \cellcolor{mygray} 0.722                       & 0.594         &  &  0.650 & \cellcolor{mygray}   0.731      \\ \bottomrule[\thickline]
\end{tabular}
\end{table}

\noindent\textbf{Failure Cases.}  
\xm{We analyze that the failures often occur when the key behavioral indicators are implicitly described in input descriptions, rendering \name and vanilla LLMs ineffective.
For example, we have checked the low recalls of 0.21 (\name) and 0.0 (GPT-4o) in our own tests of Privilege-Escalation rules, the input description states ``These calls are used to spawn MSBuild.exe in a suspended state before injecting the decrypted SaintBot binary into it, modifying the thread context to point to the malicious entry point and resuming the process'' without the behavioral indicators of process hollowing- a technique often used for privilege escalation or execution evasion. Under the same inputs, \name can consistently outperform the vanilla GPT-4o.} 

\begin{tcolorbox}
\textbf{Answer to RQ1:} 
\name agent-based reasoning mechanism enhances logical structure and syntax adherence in terms of both similarity score and the execution success.
\end{tcolorbox}

\subsubsection{RQ2-Efficiency} 
We present the computational and economic costs in Table~\ref{tab:cost}, which are derived by running \name and the baseline approach on Splunk's open-source datasets (detailed in Table~\ref{tab:datasets}) and averaging the results across multiple test cases. 
We find that \name requires more tokens and computation time than the baseline approach, mainly due to its stepwise reasoning and iterative refinement.
However, this also results in more complete and logically structured rules, as seen in earlier evaluations, with accessable latency. 

\begin{tcolorbox}
\textbf{Answer to RQ2:} 
\name generates well-structured rules while maintaining accessible latency.
\end{tcolorbox}

\begin{table}[]
\footnotesize
\setlength{\abovecaptionskip}{0pt}
\setlength{\belowcaptionskip}{0pt}
\caption{Efficiency and cost of \name and baselines.}
\label{tab:cost}
\renewcommand\tabcolsep{6pt}
\centering
\setlength{\abovecaptionskip}{0pt}
\setlength{\belowcaptionskip}{0pt}
\begin{tabular}{lllllll}
\toprule
Model                        &  &           & \begin{tabular}[c]{@{}l@{}}Prompt\\ Tokens\end{tabular} & \begin{tabular}[c]{@{}l@{}}Output\\ Tokens\end{tabular} & \begin{tabular}[c]{@{}l@{}}Money\\ Cost\end{tabular} & \begin{tabular}[c]{@{}l@{}}Generation\\ Time\end{tabular} \\ \cmidrule{1-1} \cmidrule{3-7} 
\multirow{2}{*}{GPT-4o}      &  & \name &      13,752                                                   &   2489                                                      &          \$0.060                                            &            78s                                               \\
                             &  & Baseline  & 1,295                                                  & 325                                                   & \$0.012                                              & 12s                                                       \\ \cmidrule{1-1} \cmidrule{3-7} 
\multirow{2}{*}{DeepSeek-V3} &  & \name &  24,820                                                       &  4,296                                                       &    --                                                  &    158s                                                       \\
                             &  & Baseline  & 3,284                                                 & 772                                                   & --                                                   & 31s                                                       \\ \cmidrule{1-1} \cmidrule{3-7} 
\multirow{2}{*}{LLaMA-3}      &  & \name &    22,107                                                     &  2,985                                                       &                                    --               &    119s                                                       \\
                             &  & Baseline  & 1,734                                                  & 474                                                   & --                                                   & 26s                                                       \\ \bottomrule
\end{tabular}
\end{table}

\subsubsection{RQ3-Ablation Study}
\begin{figure}[htbp]
\setlength{\abovecaptionskip}{0pt}
\setlength{\belowcaptionskip}{0pt}
    \centering
\includegraphics[width=0.8\linewidth]{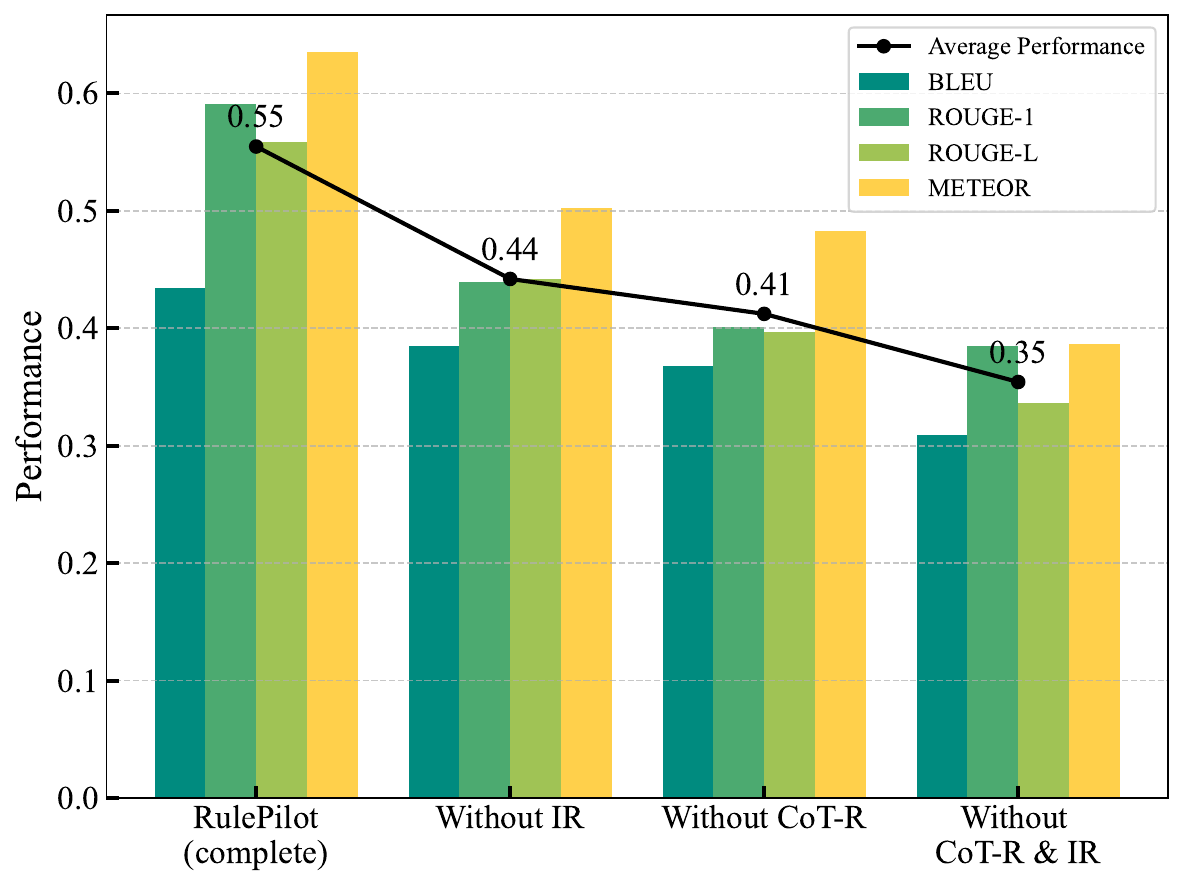}
    \caption{Ablation study on the impact of IR and CoT-R.}
    \label{fig:ablation}
\end{figure}

To further evaluate the effectiveness of the key components in \name, we conduct an ablation study focusing on two critical elements: the IR and the combination of CoT reasoning and Reflection (CoT-R). Since Reflection involves iterative refinements that call CoT modules, these two components are inherently linked and evaluated as CoT-R.
To assess the individual contributions, we introduce three experimental variants to isolate the contribution of each component: one without IR, another without CoT-R, and a version without both IR and CoT-R. The full version of \name incorporates both IR guidance and CoT-R.  
We present the results of the ablation study in Figure~\ref{fig:ablation}. The overall trend reveals that removing either IR or CoT-R leads to a significant decrease in rule generation, and removing both components causes the most substantial drop across all metrics. 
Without CoT-R and IR, the model struggles to handle complex conditions and multi-step logic, leading to incomplete or logically inconsistent rules. 
This suggests that CoT-R and IR play a critical role in enabling the model to break down complex rule-generation tasks into manageable steps, resulting in better logical consistency and structural coherence.

\begin{figure}[htbp]
\setlength{\abovecaptionskip}{0pt}
\setlength{\belowcaptionskip}{0pt}
    \centering
\includegraphics[width=0.85\linewidth]{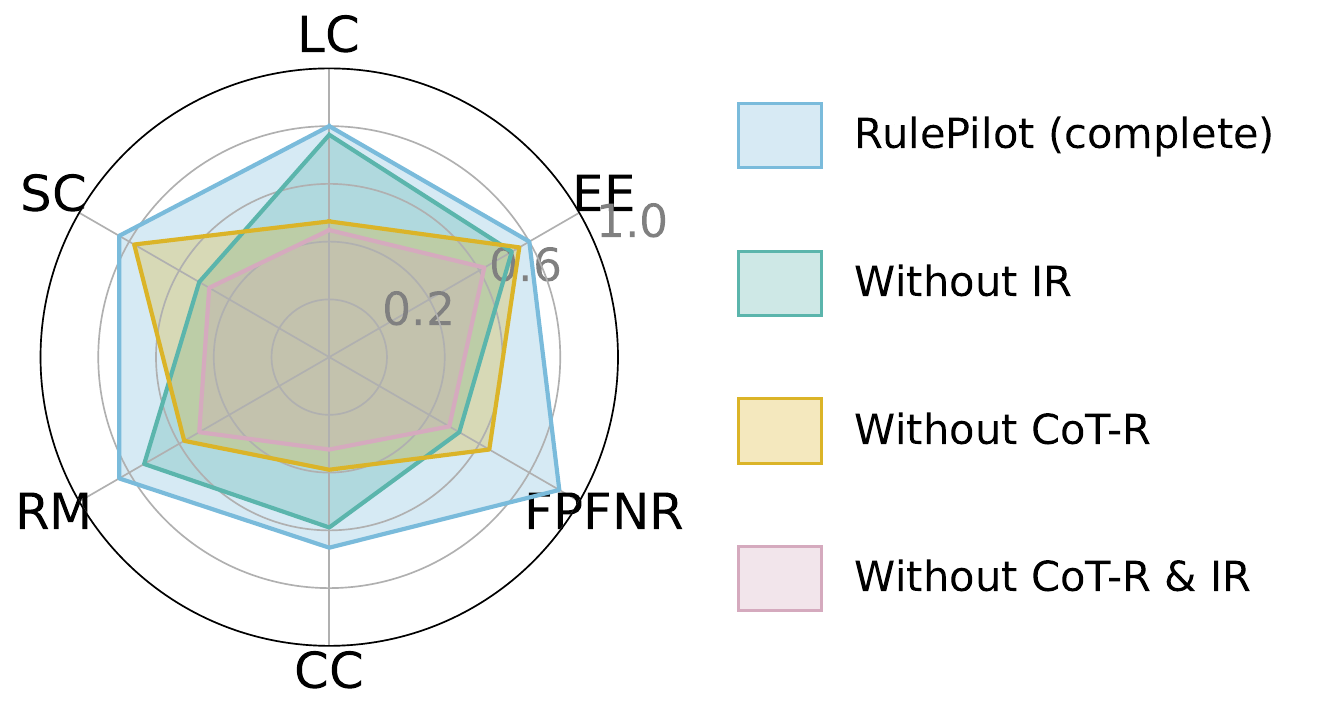}
    \caption{Ablation study for semantic-level evaluation between \name and its variants.}
    \label{fig:ablation-radar}
\end{figure}

Additionally, to determine the specific areas influenced by IR and CoT-R, we conduct a semantic evaluation, with results shown in Figure~\ref{fig:ablation-radar}. 
We find that removing CoT-R causes the most significant degradation in Logical Consistency (LC) and Condition Coverage (CC).
Conversely, removing IR primarily affects Syntax Correctness (SC) and Readability \& Maintainability (RM). 
The findings further highlight the complementary roles of these components, where CoT-R enhances logical structuring, and IR ensures syntactic correctness and standardization.

\begin{tcolorbox}
\textbf{Answer to RQ3:} Both IR and CoT-R improve rule generation. CoT-R helps with logic and structuring, while IR ensures correct syntax and readability. Removing either one lowers performance, and removing both causes the biggest drop. 
\end{tcolorbox}

%% file: 4-RQ4.tex
\subsection{RQ4-Compatibility} 
To evaluate the compatibility of \name, we use the dataset consisting of 30 SPL rules and their corresponding 30
KQL rules, which serve as ground truth references. 
These rules are sourced from real-world security applications within our industry collaborator, having been collected, segmented, and anonymized to eliminate sensitive information while preserving their syntactic and structural integrity.  
We further categorize the dataset into three types: aggregation-based rules, which summarize event data; list-based rules, which group multiple attributes; and join-based rules, which are complex and involve multi-source data correlation to detect cross-event security patterns. 
Each category contains 10 pairs of SPL and KQL rules. 
\wht{To maintain consistency, we collected logs oriented to Splunk, and evaluate the execution success on Splunk based on cases of converting KQL to SPL.} 
For each rule, we convert the given KQL query into an SPL query using our model, execute it in Splunk, and compare its results with the original SPL query from the dataset.
If both queries retrieve the same logs under identical conditions, the conversion is considered successful. 

\noindent \textbf{Evaluation Results.} We present the evaluation results in Table~\ref{tab:conversion-result} for 
rule conversion, categorized by rule type. The aggregation-based and list-based rules achieve perfect precision, recall and F1 (1.000), indicating that these rule types are straightforward to convert due to their simple structure and direct function mappings between SPL and KQL. Since they primarily involve statistical summarization or attribute grouping, the model can accurately translate their logic without ambiguity. 
However, join-based rules exhibit slightly lower performance. 
These rules involve multi-source data correlation, requiring careful field mapping and handling of log relationships across different event sources. The drop in performance is primarily due to boundary cases where certain event correlation logic was not fully preserved, leading to minor mismatches in retrieved log sets. Despite this, the results demonstrate that the conversion model is highly effective across different rule types, particularly for structured and statistical queries.

\begin{table}[htbp]
\footnotesize
\setlength{\abovecaptionskip}{0pt}
\setlength{\belowcaptionskip}{0pt}
\centering
\caption{Evaluation of rule conversion \wht{from KQL to SPL.}} 
\label{tab:conversion-result}
\renewcommand\tabcolsep{8.5pt}
\begin{tabular}{llccc}
\toprule
Rule Type               &  & Precision ($\uparrow$) & Recall ($\uparrow$) & F1 ($\uparrow$) \\ \cmidrule{1-1} \cmidrule{3-5} 
Aggregation-Based Rules &  &  1.000         &  1.000      &   1.000          \\
List-Based Rules        &  &  1.000         &  1.000      &   1.000          \\
Join-Based Rules        &  &  0.926         &   0.913     &  0.919          \\ \bottomrule
\end{tabular}
\end{table}

\xm{This experiment demonstrates the compatibility and generalization capability of \name for cross-SIEM rule conversion. While this experiment focuses on KQL-to-SPL translation due to log availability constraints (i.e., we collected logs oriented to Splunk SIEMs for execution success), \name is inherently designed to support flexible and bidirectional conversions across multiple SIEM platforms.
SPL2KQL\footnote{https://azure.github.io/spl2kql/dist/index.html} is one of the publicly available rule conversion tools used in industry. Developed by Microsoft, it supports one-way translation from Splunk SPL to Microsoft Sentinel's KQL. SPL2KQL is primarily designed to ingest external detection rules into the Microsoft ecosystem and is based on traditional rule rewriting techniques such as keyword mapping, syntax tree parsing, and regex-based transformation. However, it does not support reverse conversion or semantic adaptation for other platforms. 
In contrast, \name leverages LLM-based semantic understanding and an intermediate representation (IR) layer to support \textit{bidirectional and context-aware rule conversion}, such as KQL-to-SPL, SPL-to-KQL, or even translation between other vendor formats. This flexibility makes \name applicable to a wider range of deployment scenarios, including hybrid or transitioning security infrastructures.}

\begin{figure}[htbp]
\setlength{\abovecaptionskip}{0pt}
\setlength{\belowcaptionskip}{0pt}
    \centering
\includegraphics[width=\linewidth]{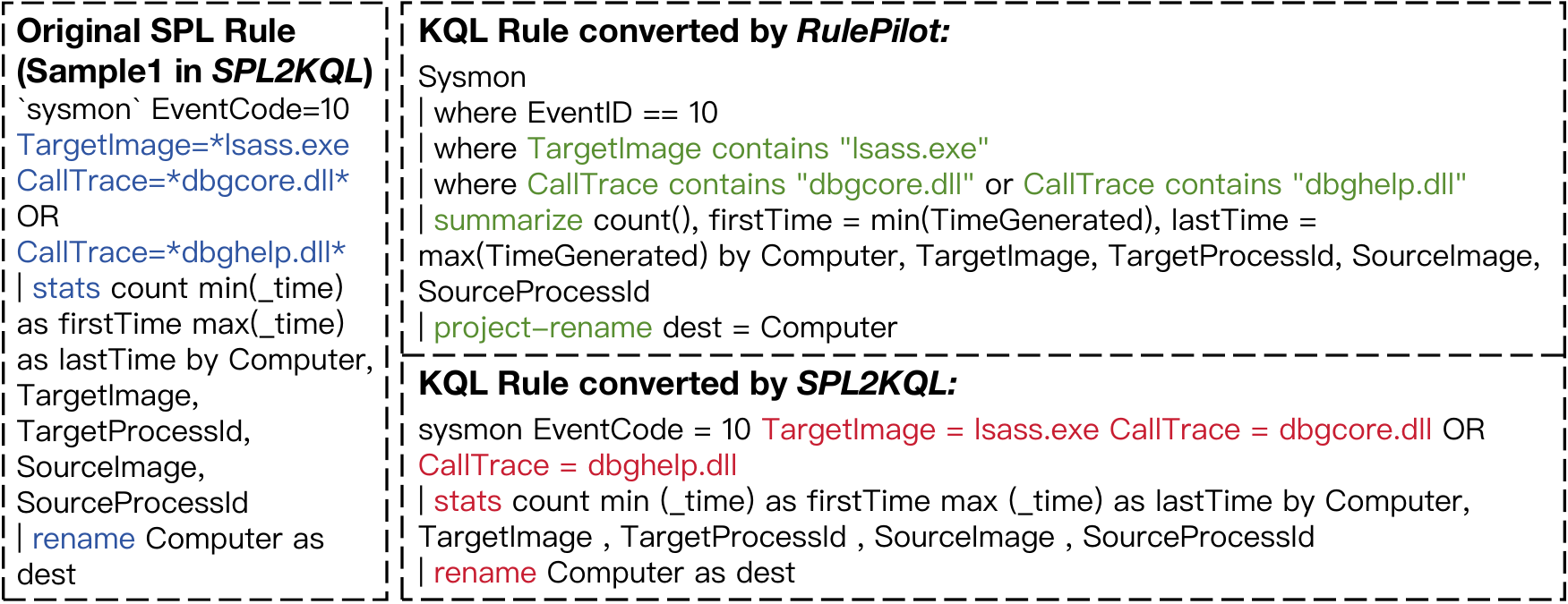}
    \caption{\xm{Comparison between the KQL rules converted from SPL via SPL2KQL and \name.}}
    \label{fig:rulepilot-vs-spl2kql}
\end{figure}

\xm{To provide a more intuitive comparison when converting SPL to KQL, we select one piece of SPL from the official SPL2KQL demo repository and convert the SPL to KQL using both SPL2KQL and \name.
As shown in Figure~\ref{fig:rulepilot-vs-spl2kql}, \name can generate a semantically faithful KQL rule by aligning query operators (e.g., \colorbox{mygray}{\texttt{contains}}, \colorbox{mygray}{\texttt{project-rename}}) and adapting field references such as \colorbox{mygray}{\texttt{TimeGenerated}}, reflecting a deep understanding of both source and target semantics. In contrast, SPL2KQL applies literal keyword mappings (e.g., \colorbox{mygray}{\texttt{TargetImage = lsass.exe}}) and syntactic transformations (e.g., \colorbox{mygray}{\texttt{rename}}) without semantic reinterpretation, resulting in inaccurate or even invalid KQL logic in practical use.}

\begin{tcolorbox}
\textbf{Answer to RQ4:} 
\name effectively supports rule conversion between Splunk SPL to Microsoft KQL, supporting the abilities of translating multiple types of rules across SIEM systems. 
\end{tcolorbox} 

%% file: 6-discussion.tex
\subsection{Case Study}
\xm{We perform a case study to compare the statistical labor reduction using \name, assessing how users of different security expertise levels perform in rule authoring with and without its support.} 
\xm{We recruit \textbf{general users} without any background of SIEM environments and the \textbf{junior analysts} with beginner experience with SIEM exposure, under the premise that \name incorporates expert-level expertise. We evaluate the time taken (Time used to produce a complete rule), final rule output, syntax validity (whether the rule passes vendor-side syntax checks, e.g., Splunk), and logical alignment (whether the rule logic matches the input as judged by an expert). The details are shown in our user study in ~\footnote{\url{https://sites.google.com/view/rulepilot/user-study}.}.}

\xm{The comparison study show that \name can significantly improve the manul rule generation process for both both general users and junior analysts, reducing the time required and improving rule quality in terms of syntactic validity and logical alignment with expert-level standards.}









\section{Discussion and Related Works}

\noindent\textbf{Automation Level.}
\name achieves a half-automated approach to SIEM-specific rule generation by embedding the logic and expertise of senior analysts. It simulates their decision-making process, including pipeline breakdown, formal template structuring, and iterative refinement. However, in practice, certain field validations and the final results require human oversights, which junior experts can handle to ensure functional-correctness and reliability.    
The operator is expected to be familiar with the SIEM environments. 
Compared to manual rule creation, the junior experts here focus on validation, eliminating the need to master complex rule grammars.

\noindent\textbf{Constraint Generation.}
Recent studies have utilized LLMs to generate constraint logic rules in various domains~\cite{ConcoLLMic-agent, Luo-protocol-fuzzing, DBLP:conf/ccs/XuWYZZH21:chunk}. For instance, LLMs have been applied to formal verification tasks in smart contracts~\cite{DBLP:journals/corr/abs-2405-02580:propertyGPT}, and to the automated extraction of generic-signature detection rule candidates from textual and visual open-source cyber threat intelligence data~\cite{schwartz2024llmcloudhunter}. 
Additionally, LLMs have been explored for log-based anomaly detection~\cite{qi2023loggpt}, demonstrating the potential of LLMs in leveraging pre-trained knowledge to extract structured insights from large-scale log data and assist in constraint generation.
However, challenges persist in modeling and capturing the intricate structures of SIEM rules, hindering the direct application of these methods to generate executable security rules.

\noindent\textbf{Log Analysis.}  
Previous works largely employ LLMs to automate log analysis~\cite{liu2024interpretable, qi2023loggpt, }, including log parsing and anomaly detection. For instance, LLM-based approaches achieve high precision in log template extraction~\cite{xu2024divlog} and automatic logging statement generation~\cite{xu2024unilog}, significantly reducing manual effort. 
These approaches may provide valuable foundations for our work by improving log parsing and structured analysis.
Unlike prior studies that focus on general log processing, our work builds upon existing SIEM rules and leverages LLMs to analyze logs and detect anomalies.



\section{Conclusion}
In this paper, we propose \name, an LLM-based agent system designed to automate rule creation and conversion for SIEM detection. By leveraging \xm{the novel SIEM-specific} intermediate representation, \name abstracts the complexity of rule configurations into a structured and standardized format. 
We conduct a comprehensive evaluation of \name, demonstrating that it can produce high-fidelity, executable SIEM-specific rules.  
\wht{Our case study with industry collaborators shows that \name significantly assists general users and junior analysts by reducing rule generation time and improving rule quality, allowing them to create detection logic using natural language instead of manually adhering to strict grammar rules.}

\begin{acks}
We thank the anonymous meta review and all anonymous reviewers for their insightful comments to improve this paper. This paper is supported by NUS-NCS Joint Laboratory for Cyber Security. 
\end{acks}